\documentclass[10pt]{llncs}
\usepackage{amsmath, amssymb, algorithm, algorithmic, color}
\usepackage{enumerate, graphicx, hyperref}
\graphicspath{{figures/}}
\usepackage{cite}
\usepackage{todonotes}
\usepackage{microtype}

\let\doendproof\endproof
\renewcommand\endproof{~\hfill\qed\doendproof}

\newcommand{\nonterm}[1]{\langle\text{#1}\rangle}
\newcommand{\term}[1]{\text{#1}}

\title{Confluent Orthogonal Drawings\\of Syntax Diagrams}
\author{Michael J. Bannister
\and David A. Brown
\and David Eppstein
}

\institute{Department of Computer Science, University of California, Irvine\thanks{Michael Bannister and David Eppstein were supported in part by NSF grant  CCF-1228639.}}

\begin{document}
\maketitle

\begin{abstract}
We provide a pipeline for generating syntax diagrams (also called railroad diagrams) from context free grammars. Syntax diagrams are a graphical representation of a context free language, which we formalize abstractly as a set of mutually recursive nondeterministic finite automata and draw by combining elements from the confluent drawing, layered drawing, and smooth orthogonal drawing styles. Within our pipeline we introduce several heuristics that modify the grammar but preserve the language, improving  the aesthetics of the final drawing.
\end{abstract}

\section{Introduction}
The languages of computing, such as programming languages and data exchange formats, are typically specified using a finite set of rules called a grammar, and these rules are usually given in Backus--Naur Form or one of its extensions. Backus--Naur Form provides a notation rich enough to express all context-free grammars, and in turn most grammars of practical interest, while being easily machine readable. However, being a purely textual representation, it is perhaps less  readable by humans. For this reason, Jensen and Wirth used a graphical representation of context-free grammars, called syntax diagrams, when defining the programming language Pascal~\cite{pascal-book}.\footnote{Jensen and Wirth were not the first to use syntax diagrams~\cite{cande-book}, but they popularized them, and these diagrams have been widely used since.} We investigate the problem of generating syntax diagrams for context-free grammars and provide several heuristics optimizing the aesthetics of the resulting drawing. Our work provides the first algorithmic study of this problem and the first system that attempts to optimize the resulting diagram for readability rather than directly translating a given grammar into a diagram.

Recall that a \emph{context-free grammar} is defined by four values $N,\Sigma,R,S$. In this $4$-tuple, $N$ is a set of \emph{nonterminal symbols}, $\Sigma$ is a set of \emph{terminal symbols}, $R$ is a set of \emph{production rules} of the form $A \to \beta$ where $A$ is a nonterminal symbol and $\beta$ is a (possibly empty) string of terminal and nonterminal symbols, and $S$ is a nonterminal symbol designated as the \emph{start symbol}. A string $\sigma$ of terminal symbols belongs to the language defined by the grammar when there exists a sequence of \emph{rewrite steps} starting from $S$ and ending at $\sigma$, each of which replaces a nonterminal symbol $A$ in the current string with a string $\beta$ such that $A \to \beta$ is a production rule in the grammar. \autoref{fig:s-expression-grammar} gives an example grammar for the S-expressions in the programming language LISP~1.5.
\begin{table}[t]
\begin{align*}
\nonterm{S-expression} &\to \nonterm{atomic symbol}\\
                       &\quad\mid  (\nonterm{S-expression} . \nonterm{S-expression})\\
                       &\quad\mid (\nonterm{S-expression list})\\
\nonterm{S-expression list} &\to \epsilon \mid \nonterm{S-expression} \nonterm{S-expression list}\\
\nonterm{atomic-symbol} &\to \nonterm{LETTER}\nonterm{atom part}\\
\nonterm{atom part} &\to \epsilon \mid \nonterm{LETTER}\nonterm{atom part} \mid \nonterm{number}\nonterm{atom part}\\
\nonterm{LETTER} &\to \term{A} \mid \term{B} \mid \term{C} \mid \cdots \mid \term{Z}\\
\nonterm{number} &\to \term{0} \mid \term{1} \mid \term{2} \mid \cdots \mid \term{9}
\end{align*}
\caption{A context-free grammar for the language of S-expressions in LISP~1.5~\cite{lisp-book}.}
\label{fig:s-expression-grammar}
\end{table}

A \emph{regular grammar} is one in which the production rules all have the form $A \to b$, $A \to bC$ or $A \to \epsilon$, where $A$ and $C$ are nonterminals, $b$ is a terminal, and $\epsilon$ is the empty string. An example of a regular grammar is the part of the LISP~1.5 grammar defining $\nonterm{atom part}$.
%
%An example of a regular grammar for the $\nonterm{atom part}$ of S-expressions is given in \autoref{fig:atom-part-grammar}.
%\begin{figure}[t]
%\begin{align*}
%\nonterm{atom part} &\to \epsilon
%                    \mid \term{A}\nonterm{atom part} \mid \term{B}\nonterm{atom part} \mid \cdots \mid \term{Z}\nonterm{atom part}\\
%&\hspace{2em}\mid \term{0}\nonterm{atom part} \mid \term{1}\nonterm{atom part} \mid \cdots \mid \term{9}\nonterm{atom part}
%\end{align*}
%\caption{A regular grammar for the $\nonterm{atom part}$ part of S-expressions.}
%\label{fig:atom-part-grammar}
%\end{figure}
%
Languages definable by regular grammars are exactly the regular languages, whose equivalent characterizations include being recognizable by nondeterministic finite automata (NFAs). For these languages, we could use graph drawings of an NFA state graph as a graphical representation, by drawing an $st$-digraph with edges labeled by terminal symbols. A string $\sigma$ is in the language if and only if there is a directed path through the graph from $s$ to $t$ such that the concatenation of the edge labels is equal to $\sigma$. Unfortunately, such a representation will not work for non-regular languages.

To graphically represent context-free languages we turn to syntax diagrams.
Although other authors used syntax diagrams earlier~\cite{cande-book}, they were popularized by the Pascal User Manual and Report by Jensen and Wirth~\cite{pascal-book}. The style has been praised for its readability~\cite{Bra-SJCD-1990} and pedagogical value~\cite{BelGil-Bull-1974}, and has been used by the Smalltalk-80 Blue Book~\cite{smalltalk-book}, JSON Data Interchange Standard~\cite{json-web}, and the W3C technical report on CSS~\cite{css-web}. Several software packages have been created to automate the drawing of syntax diagrams~\cite{sw1-web,sw2-web,sw3-web}. These software packages provide little to no optimization of the drawing, providing only a one-to-one translation of the Extended Backus--Naur grammars into syntax diagrams. Until now, there does not seem to be any algorithmic research involving the generation and optimization of syntax diagrams.

We introduce a new formalization for syntax diagrams consisting of a collection of $st$-digraphs (see e.g., \autoref{fig:nfa-representation-lisp}), each representing the possible expansions of a single nonterminal symbol, with each edge in each graph labeled by either a terminal or a nonterminal symbol. As before a string is in the language if and only if the string can be represented by a directed path from $s$ to $t$ in the start symbol's $st$-digraph. However, when this path would contain a nonterminal symbol, we recurse into the $st$-digraph corresponding to that symbol. The concatenation of the terminal symbols in the resulting system of recursively generated paths should match the sequence of terminal symbols in the given string.

\begin{figure}
\centering
\includegraphics[width=0.49\textwidth]{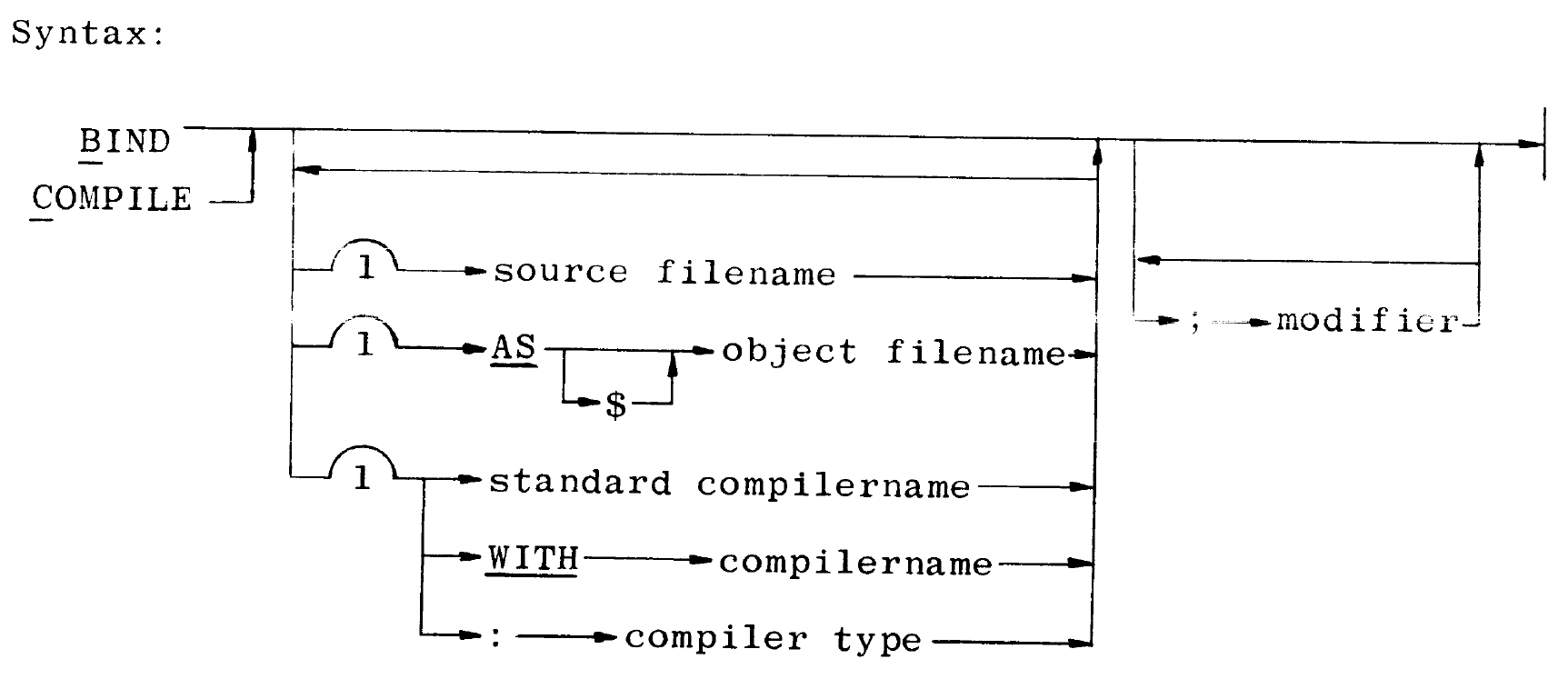}
\hfill
\includegraphics[width=0.49\textwidth]{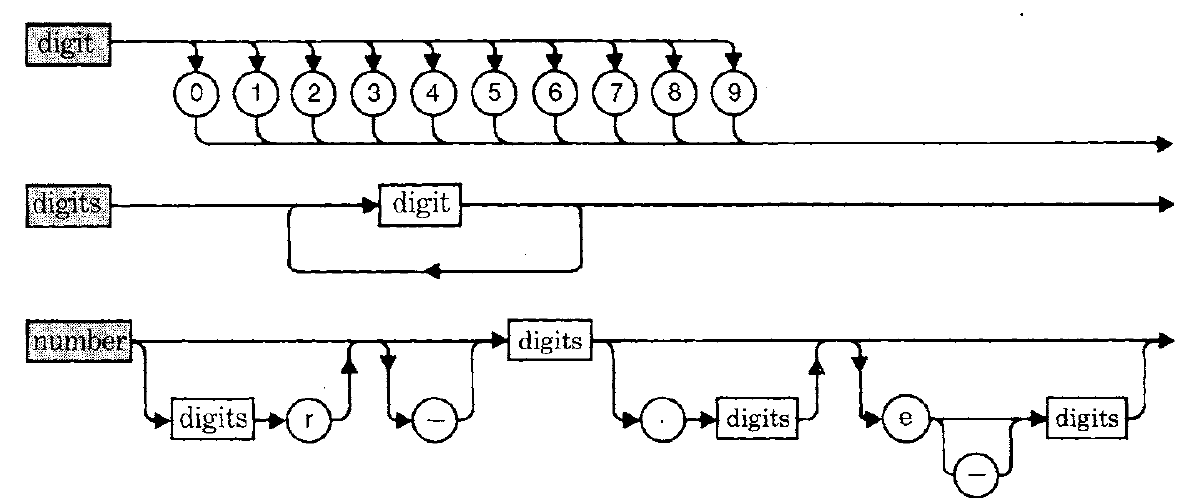}
\caption{A syntax diagram from the CANDE Information Manual (left) and a confluent syntax diagram from the Pascal User Manual and Report (right).}
\label{fig:s-expressions-syntax-diagram}
\end{figure}

Without further optimization this formalization merely gives a new notation for writing production rules, but it has two advantages over extended BNF. Firstly, it gives us additional freedom in our representation: a BNF grammar can only describe syntax diagrams formed by a collection of disjoint paths between the two terminals, and extended BNF can still only describe syntax diagrams in the form of series-parallel graphs, while our diagrams are not restricted in these ways. Secondly, as we describe below, we can use this notation to directly represent the junctions and tracks of a confluent drawing style~\cite{DicEppGooMen-GD-04}, in which a path through the graph is only valid if it is a smooth path, such as in \autoref{fig:s-expressions-syntax-diagram} (right). It is this drawing style that gives rise to the occasionally used alternative name ``railroad diagrams'' for syntax diagrams.

Our drawings will combine confluent drawing with Sugiyama-style layered drawing~\cite{SugTagSho-SMC-1981,BasMat-DGMM-01} using smooth orthogonal edge shapes~\cite{BekKauKob-JGAA-13}. The combination of confluent and layered drawing has been studied before~\cite{EppGooMen-GD-05}, but in a different way. Past work considered confluent drawing as a technique for visualizing a specific graph, and involved a search for subgraphs that could be more concisely expressed using confluence. In our application, the graph (NFA) representation that we work with already encodes the confluent features of the drawing: its vertices become confluent junctions in the drawing, and its edges become the boxes and connecting segments of track of the drawing (\autoref{fig:confluent-conversion}). Rather than searching for graph features that can become confluent, our focus is on modifying the underlying NFA to produce a simpler and higher-quality drawing while preserving the equivalence of the underlying context-free language described by the drawing.

\subsection{Software pipeline}
\begin{figure}[t]
\includegraphics[width=\textwidth]{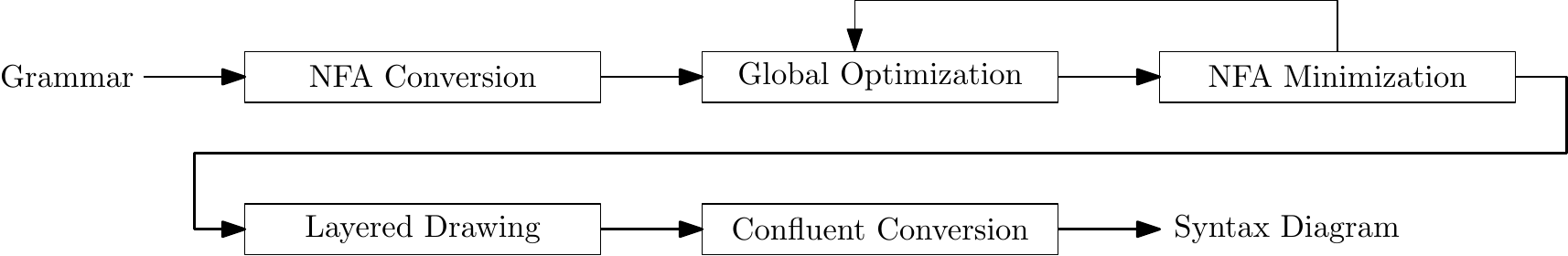}
\caption{A flow chart describing our software pipeline.}
\end{figure}
\begin{figure}
\centering
\includegraphics[scale=0.65]{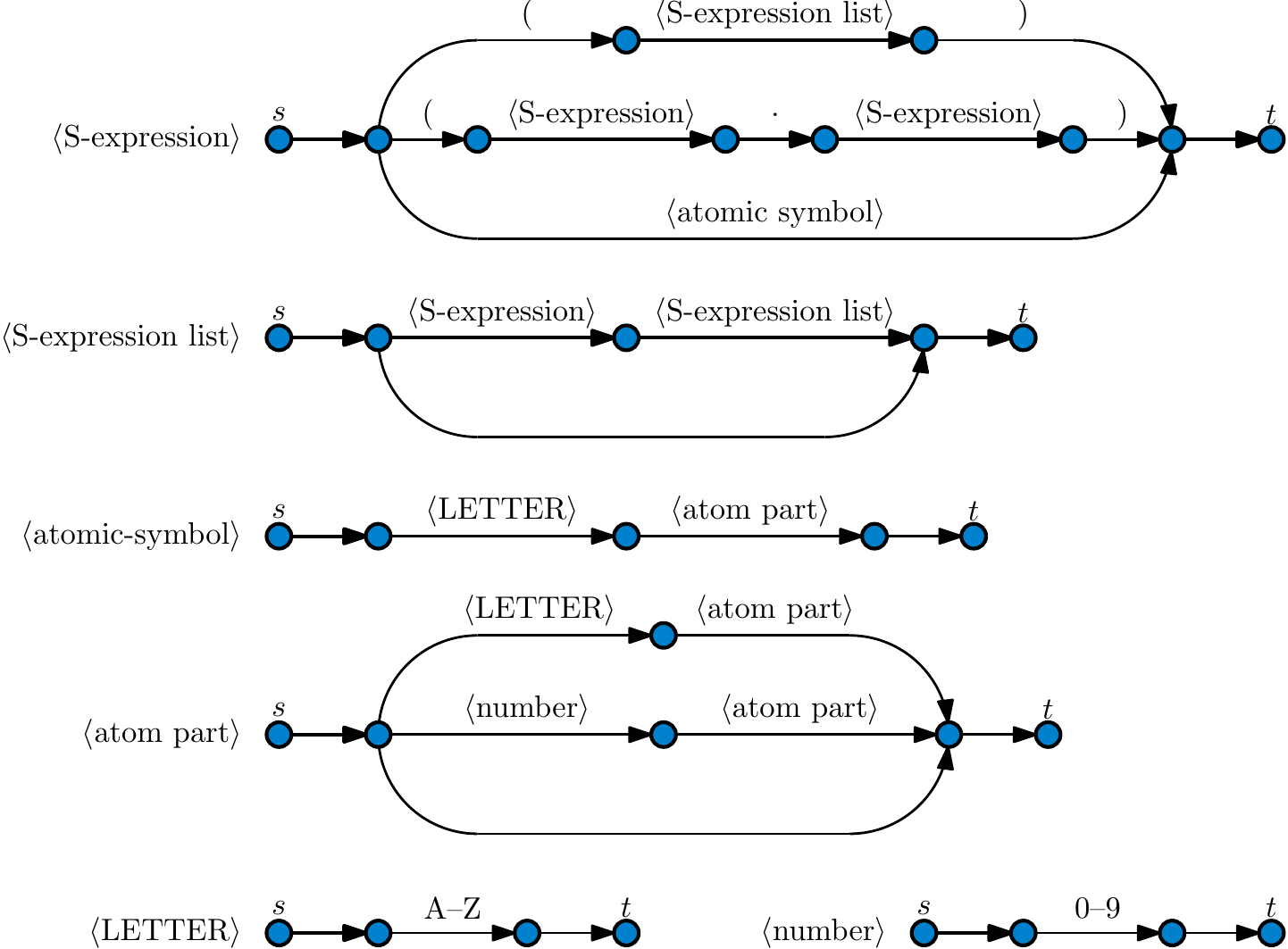}
\caption{The initial NFA-representation of S-expressions in the LISP~1.5 grammar.}
\label{fig:nfa-representation-lisp}
\end{figure}

We describe our method for producing syntax diagrams with the framework of a generic software pipeline. In the first step of our pipeline, we convert the grammar to our internal representation, which we will call the \emph{NFA representation}. This representation consists of a family of $st$-digraphs, initially one for each nonterminal symbol, whose edges are labeled by (terminal and nonterminal) symbols in the grammar or $\epsilon$ (the empty string). To construct the $st$-digraph for the nonterminal symbol $A$ we convert each production of the form $A \to B_0B_1\cdots B_{r-1}$ into a directed path of length $r$ labeled by the symbols $B_0, B_1, B_{r-1}$. Then all of the beginning and ending vertices are respectively merged together. Finally, we add to the graph two extra $\epsilon$-labeled edges, one at the beginning and one at the end. See \autoref{fig:nfa-representation-lisp} for the complete NFA representation of LISP~1.5.

The second and third steps in the pipeline attempt to reduce the number of total symbols in the NFA representation, through both global optimizations that act on the entire system of graphs and local optimizations that act on a single graph. The local optimization part of the pipeline is a form of the well-studied problem of NFA minimization. In general exact NFA minimization is $\mathsf{PSPACE}$-hard~\cite{HunRosSzy-JCSS-1976, StoMey-STOC-1973}, and furthermore approximating the minimum NFA efficiently to within an $o(n)$ approximation ratio is also $\mathsf{PSPACE}$-hard~\cite{GraSch-JCSS--2007}. However, since the problem is of practical importance there are many heuristic approaches~\cite{ChaCou-TCS-2004, HanWoo-TCS-2007}. In this paper, we use simple heuristics motivated by the structure of real-world grammars and typical simplifications found in hand drawn syntax diagrams, rather than attempting to implement the more complex heuristics devised for minimizing NFAs without regard to their appearance as a diagram.

Once the NFA representation is optimized, we draw each of the $st$-digraphs in a layered Sugiyama style~\cite{SugTagSho-SMC-1981,BasMat-DGMM-01}, rotated horizontally to direct edges from left to right. In these graphs, the only directed cycles come from tail recursion elimination, so rather than searching for a small feedback arc set to determine the reversed edges in the drawing, we maintain such a set during the process of NFA minimization and add to it whenever we perform a tail recursion elimination step. In this way, we can ensure that all the tokens in the drawing are traversed from left to right. Standard layered drawing optimizations are applicable in this stage, but were not implemented in our experiments as we were primarily interested in optimizing the NFA representation. Finally, we convert the layered drawing into a confluent syntax diagram.

\subsection{Contributions}
Our contributions in this paper are summarized below.
\begin{itemize}
\item We formalize an abstract representation of syntax diagrams as a collection of mutually recursive NFAs, allowing the application of NFA minimization heuristics beyond what is possible with EBNF.
\item We formulate a software pipeline for producing syntax diagrams, based on NFA minimization and confluent layered graph drawing.
\item We develop a family of fast and simple NFA minimization heuristics, together with global heuristics that recombine multiple NFAs.
\item We describe a geometric layout method based on a horizontal Sugiyama layered drawing, where we reinterpret the vertices and edges in a layered drawing of an NFA as the junctions and vertices of a confluent drawing.
\item We provide a proof-of-concept implementation that produces human quality syntax diagrams for real-world context-free languages.
\item Finally, we experimentally evaluate the quality of our heuristics.
\end{itemize}

\section{Global minimization heuristics}
\begin{figure}
\centering
\includegraphics[scale=0.65]{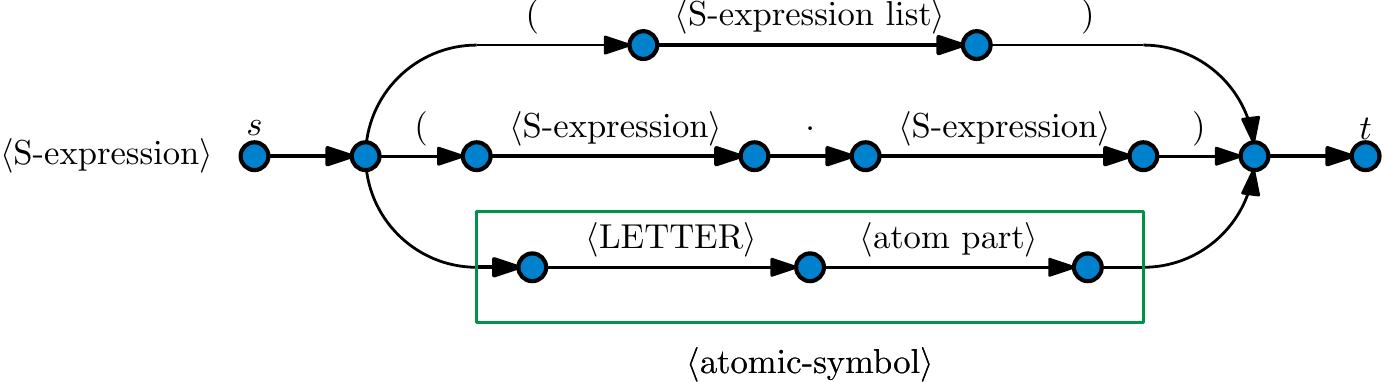}
\caption{An example of nesting the $\nonterm{atomic symbol}$ $st$-digraph into the $\nonterm{S-expression}$ $st$-digraph, within the LISP~1.5 grammar.}
\label{fig:nesting-lisp}
\end{figure}

A \emph{global minimization heuristic} seeks to minimize the total number of labeled edges in an NFA representation via the modification of two or more of the $st$-digraphs in the representation. The only global heuristic that we consider is \emph{nonterminal nesting}, in which a single nonterminal edge in one graph is replaced by the entire graph corresponding to that nonterminal edge. Since the goal is to reduce the total number of symbols in the NFA representation, we enforce the following restrictions when nesting a graph $H$ (corresponding to a nonterminal~$A$) into another graph $G$:
\begin{itemize}
\item $A$ cannot be the start symbol.
\item $G$ and $H$ must be two distinct graphs.
\item If $H$ has more than one non-$\epsilon$ edge, then $A$ must occur only once in the whole system of digraphs, and its occurrence must be in $G$.
\item The number of symbols in the  graph produced by nesting $H$ into $G$ must be less than a predefined threshold $k$.
\end{itemize}
The final restriction above is intended to keep the size of each individual $st$-digraph to a human-readable level. The nesting heuristic can be seen to have been used in some hand-drawn syntax diagrams (e.g., the JSON syntax diagrams), but it does not appear to be used by previous syntax diagram software. See \autoref{fig:nesting-lisp} for an example of nesting with the LISP~1.5 grammar.

\section{Local minimization heuristics}
A \emph{local minimization heuristic} seeks to minimize the total number of labeled edges in a single $st$-digraph within the NFA representation. Many of these optimizations can be seen in hand-drawn syntax diagrams.

\subsection{Tail recursion loop back}
\begin{figure}
\centering
\includegraphics[scale=0.65]{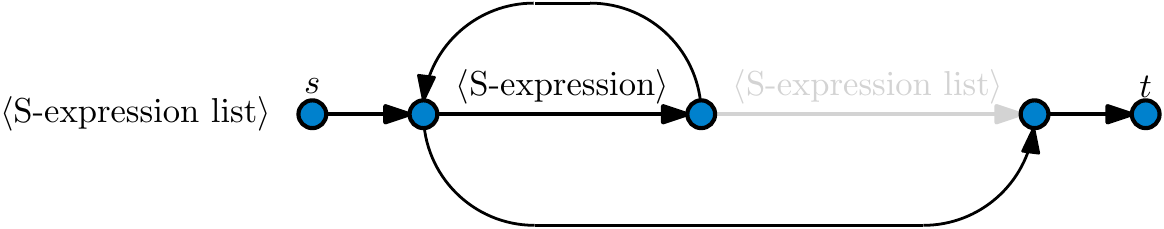}
\caption{An example of tail recursion loop back of $\nonterm{S-expression list}$ in the LISP~1.5 grammar. The removed edge has been colored gray.}
\label{fig:loopback-lisp}
\end{figure}

The $st$-digraphs produced from a grammar, before optimization, are acyclic, and nesting preserves acyclicity. However, hand-drawn syntax diagrams typically contain cycles, which we introduce as a replacement for tail-recursive grammars using the \emph{loop back heuristic}. If a nonterminal $A$ appears exactly once in its own $st$-digraph and the edge on which it appears has $t'$ (the only incoming neighbor of $t$) as its destination, then we change the destination of the $A$-labeled edge from $t'$ to $s'$ (the only outgoing neighbor of $s$) and we change its label from~$A$ to~$\epsilon$. Although this does not reduce the number of edges in the $st$-digraph, it does reduce the number of labeled edges and improves the readability of the drawing. In addition, by reducing the number of occurrences of $A$ as a label, it may cause nesting operations to become possible that were previously forbidden. The edges that are modified by this heuristic will be the only ones directed backwards in our eventual drawings. See \autoref{fig:loopback-lisp} for an example of this construction.

\subsection{Parallel state elimination with squish heuristic}
\begin{figure}
\centering
\includegraphics[scale=0.65]{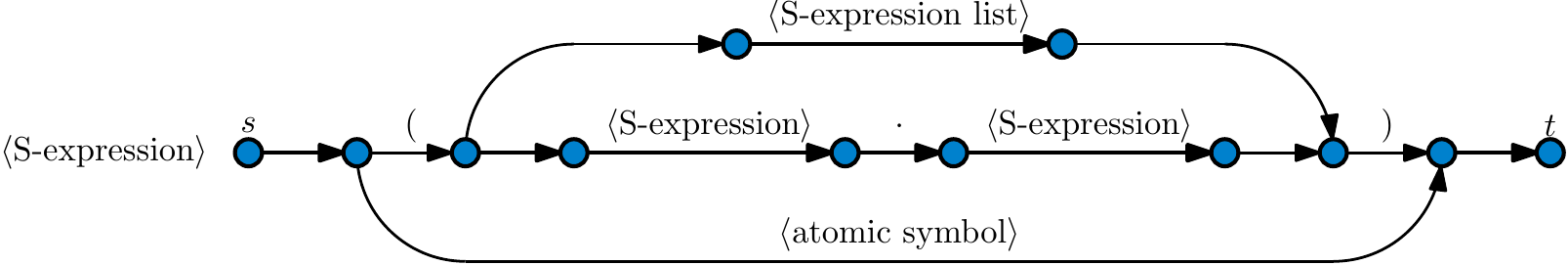}
\caption{An example of the squish heuristics applied to $\nonterm{S-expression}$ in the LISP~1.5 grammar. The squish forward combines the open parenthesis and the squish backward combines the closing parenthesis.}
\label{fig:squish-lisp}
\end{figure}

The \emph{squish forward} heuristic is used to reduce the number of nonempty symbols when there are parallel occurrences of the same symbol. If two edges $e_1=(u,v_1)$ and $e_2=(u,v_2)$ are labeled by the same symbol $A \neq \epsilon$, then we replace $e_1$ and $e_2$ with $f = (u,t)$ labeled $A$, $f_1 = (t,v_1)$ labeled $\epsilon$ and $f_2 = (t,v_2)$ labeled~$\epsilon$. We similarly define the \emph{squish backward} heuristic, to be the squish forward heuristic applied to an $st$-digraph in which all of the edges have been reversed. See \autoref{fig:squish-lisp} for an example of this heuristic.

\subsection{Epsilon transition removal}
Our previous optimizations may introduce $\epsilon$-labeled edges. We attempt to remove redundant $\epsilon$-edges using the \emph{epsilon removal} heuristic. If $e = (u,v)$, with $u\neq s$ and $v \neq t$, is an $\epsilon$ labeled edge, such that $e$ is not a reversed edge (introduced via the loop back heuristic), and either $e$ is the only outgoing edge of $u$ or the only incoming edge to $v$, then the edge $e$ is removed by merging $u$ and $v$. We iteratively find and remove such edges until no such edge exists.

\subsection{Confluent pinch}
Our final local optimization would not qualify as an NFA optimization, as it does not attempt to reduce the number of symbols. Instead, the \emph{confluent pinch} heuristic attempts to reduce crossings in the final drawing by removing directed complete bipartite subgraphs (which can be created by the squish heuristic), replacing each one by a single ``crossing'' vertex. If a digraph contains a set of vertices $U$ and a set of vertices $V$ such that there is an $\epsilon$ labeled edge $(u, v)$ for all $u \in U$ and $v \in V$, then we remove all such edges and add $\epsilon$-labeled edges $(u,w)$ for all $u \in U$ and $(w, v)$ for all $v \in V$.

\begin{figure}
\centering
\includegraphics[scale=0.65]{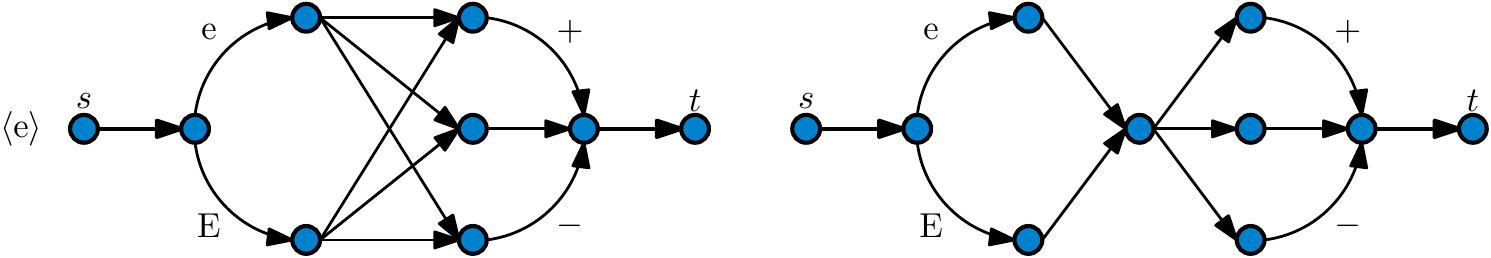}
\caption{An example of confluent pinch for scientific notation in the JSON grammar.}
\label{fig:confluent-pinch}
\end{figure}

\subsection{Implementing the heuristics}
\begin{figure}
\centering
\includegraphics[scale=0.65]{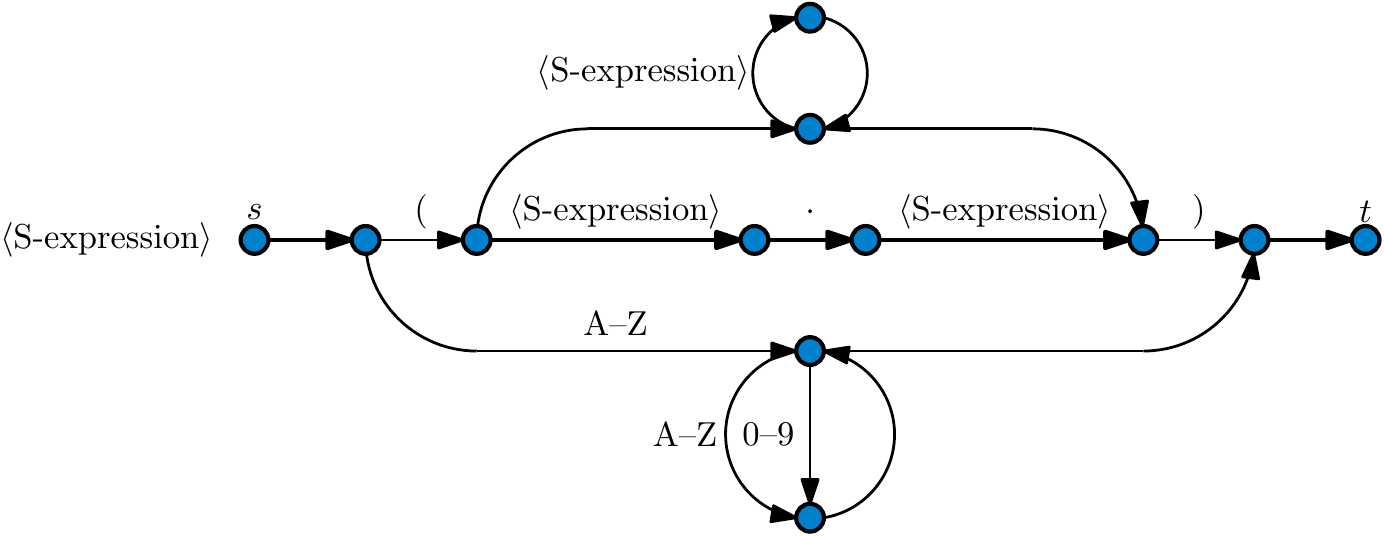}
\caption{Optimized NFA representation for S-expressions in LISP~1.5.}
\label{fig:opt-lisp}
\end{figure}

The application of one heuristic may create new optimization opportunities with respect to a previously applied heuristic. Therefore, we perform multiple rounds of optimization, applying all possible heuristics within each round, until no further optimizations are possible or a maximum number of rounds have been completed. In \autoref{fig:opt-lisp} we see the optimized NFA representation of S-expressions in LISP~1.5, as produced by our implementation of these heuristics.

\section{Sugiyama layering}
Once the NFA representation has been minimized, we give each of the $st$-digraphs a Sugiyama-style layered drawing, using the standard layered-drawing pipeline for layout and crossing minimization. One modification that we make to this pipeline is that it is neither necessary nor desirable to compute a feedback arc set of the $st$-digraphs. Instead, the set of edges introduced during the loop back heuristic already form a feedback arc set with edges which should loop back into the drawing. Since we are using an orthogonal drawing style, we add bends to edges to allow them to shift their vertical positions from one layer to the next, and use an interval-graph coloring algorithm to place the vertical connectors of these bent edges into a small number of columns.

\begin{figure}
\centering
\includegraphics[width=0.49\textwidth]{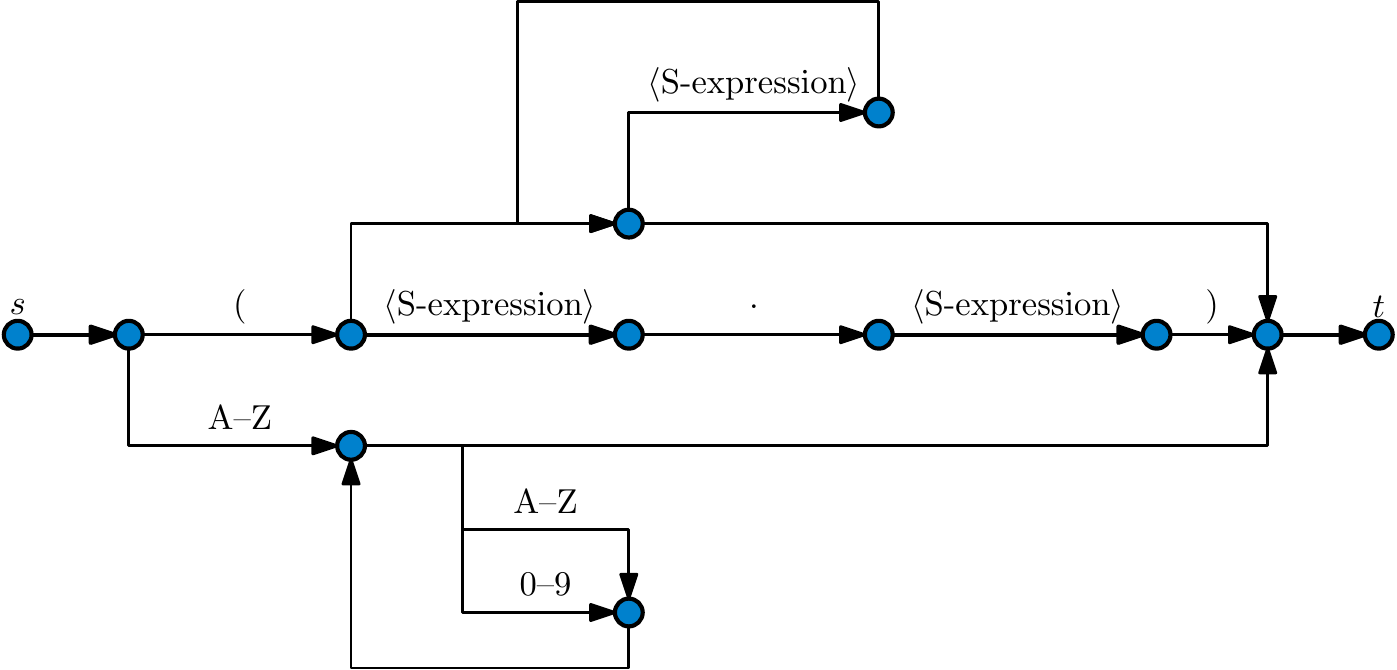}
\hfill
\includegraphics[width=0.49\textwidth]{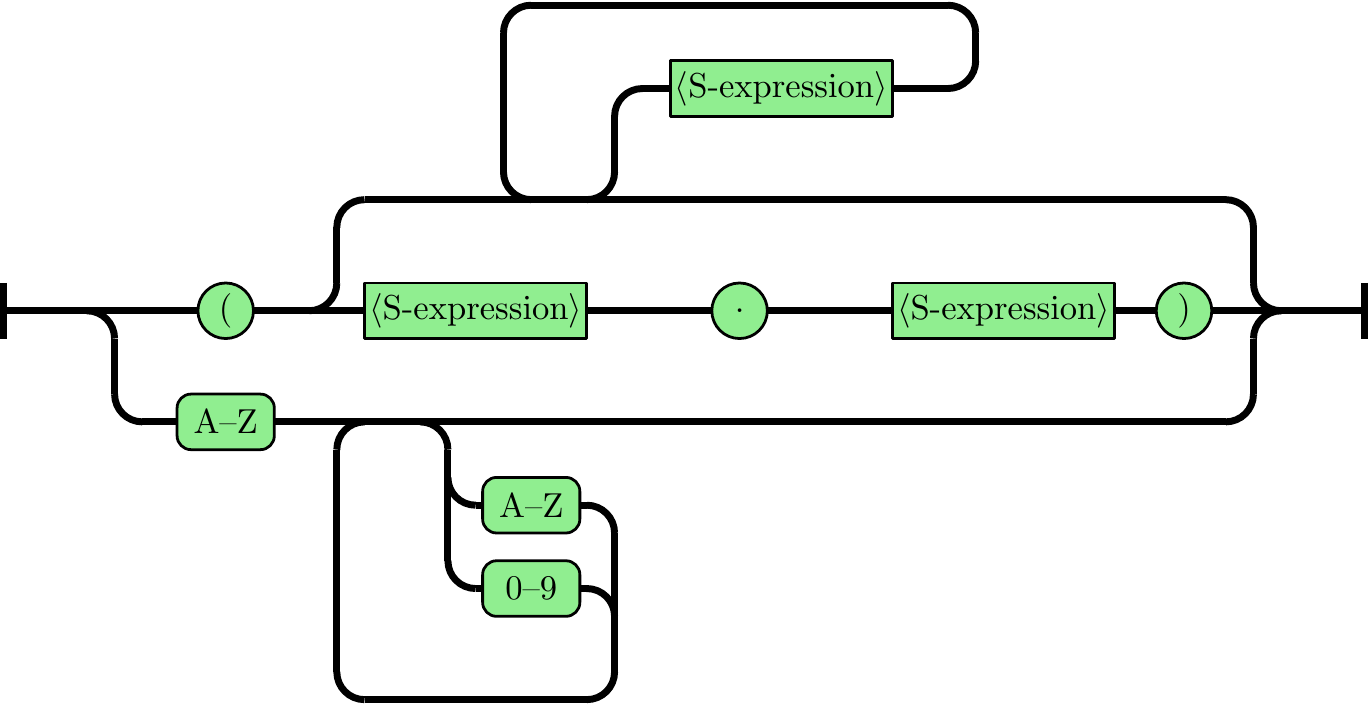}
\caption{The final confluent conversion from an orthogonal layered drawing into a syntax diagram for LISP~1.5 S-expressions.}
\label{fig:confluent-conversion}
\end{figure}

In the final step of our algorithm, we reinterpret the vertices and edges in the resulting orthogonal drawing as the confluent junctions, track segments, and vertices of a confluent drawing. We place a vertex of the confluent drawing at the middle of each edge of the layered drawing whose label is not~$\epsilon$, with the confluent vertex being given the same label as the $st$-digraph edge label.
We place a confluent junction at each vertex of the layered drawing, connected to a segment of confluent track for each incident edge of the layered drawing. Additionally, confluent junctions are created by the overlapping of edges with a common source. The orientation of the track at each confluent junction is determined by two factors: whether it connects to an earlier or a later layer, and whether it is a forward or reversed edge in the layered drawing. The result of this conversion step is our final syntax diagram. See \autoref{fig:confluent-conversion} for an example of this final conversion step.

\section{Experimental results}

In order to validate the heuristic optimizations performed by our implementation, we tested them on a set of eight real-world context-free grammars collected by Neal Wagner at the web site \url{http://www.cs.utsa.edu/~wagner/CS3723/grammar/examples2.html} together with the Lisp~1.5 and JSON grammars. For each grammar, we measured the area of our drawing (in units of rows and columns), the number of tokens (boxes) in the drawing, and the total number of connected components, both before and after optimization. The results are shown in \autoref{tbl:x}.

\begin{table}
\begin{center}
  \begin{tabular}{| l | l | c | c | c |}
    \hline
    	\textbf{Name} & \textbf{optimized?} & \textbf{area} & \textbf{tokens} & \textbf{components} \\ \hline
	Canadian post codes & unoptimized & 17 & 6 & 1 \\
	(simple) & optimized & 17 & 6 & 1 \\ \hline
	Canadian post codes & unoptimized  & 693 & 69 & 9 \\
	(complex) & optimized & 1121 & 65 & 5 \\ \hline
	Ottawa course codes & unoptimized & 520 &  46 & 15 \\
	& optimized & 570 & 36 & 5 \\ \hline
	Palindromes & unoptimized & 583 & 105 & 2 \\
	& optimized & 583 & 105 & 2 \\ \hline
	Nonempty data files & unoptimized & 182 & 22 & 8 \\
	(repetitive) & optimized  & 132 & 11 & 3 \\ \hline
	Nonempty data files & unoptimized & 143 & 22 & 7 \\
	(recursive) & optimized & 130 & 7 & 1 \\ \hline
	Pascal variable declarations & unoptimized & 156 & 21 & 7 \\
	& optimized & 247 & 12 & 3 \\ \hline
	Pascal type declarations & unoptimized & 475 & 52 & 16 \\
	& optimized & 486 & 30 & 6 \\ \hline
	LISP 1.5 & unoptimized & 165 & 19 & 6 \\
	& optimized & 105 & 9 & 1 \\ \hline
	JSON & unoptimized & 539 & 90 & 15 \\
	& optimized & 651 & 42 & 5 \\ \hline
  \end{tabular}
\end{center}
\caption{Experimental results}
\label{tbl:x}
\end{table}

As these results show, our optimizations were not always effective at reducing the total area of our drawings, and in some cases even increased the area. However, we typically achieved more significant reductions in the numbers of tokens and connected components of the drawings, which we believe to be helpful in reducing their visual clutter. Additionally, it can be seen that our optimizations are typically more effective on grammars with larger numbers of nonterminals, and less effective on grammars that have only a very small number of nonterminals, because in those cases no nesting will be possible.

We did not directly compare the results of other available syntax diagram drawing systems, but the ones we tested all appear to translate the input grammar to a diagram directly, without optimization; therefore, we believe that the results of testing them would be similar to the unoptimized lines of the table.

\section{Gallery of examples}

\renewcommand{\topfraction}{0.9}

\begin{figure}[t]
\centering
\includegraphics[width=\textwidth]{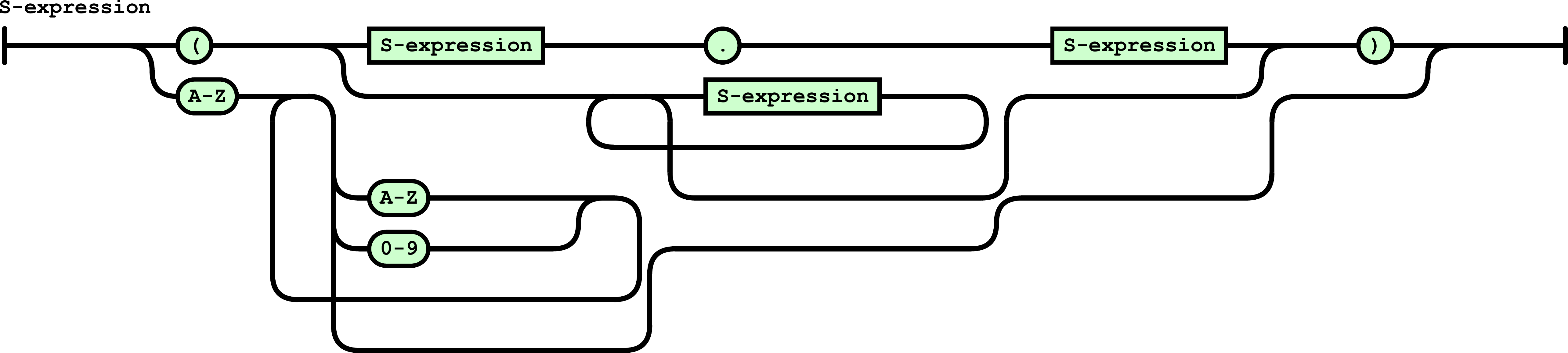}
\caption{A syntax diagram for S-expressions in LISP~1.5.}
\label{fig:complete-lisp}
\end{figure}

\begin{figure}[t]
\centering
\includegraphics[width=\textwidth]{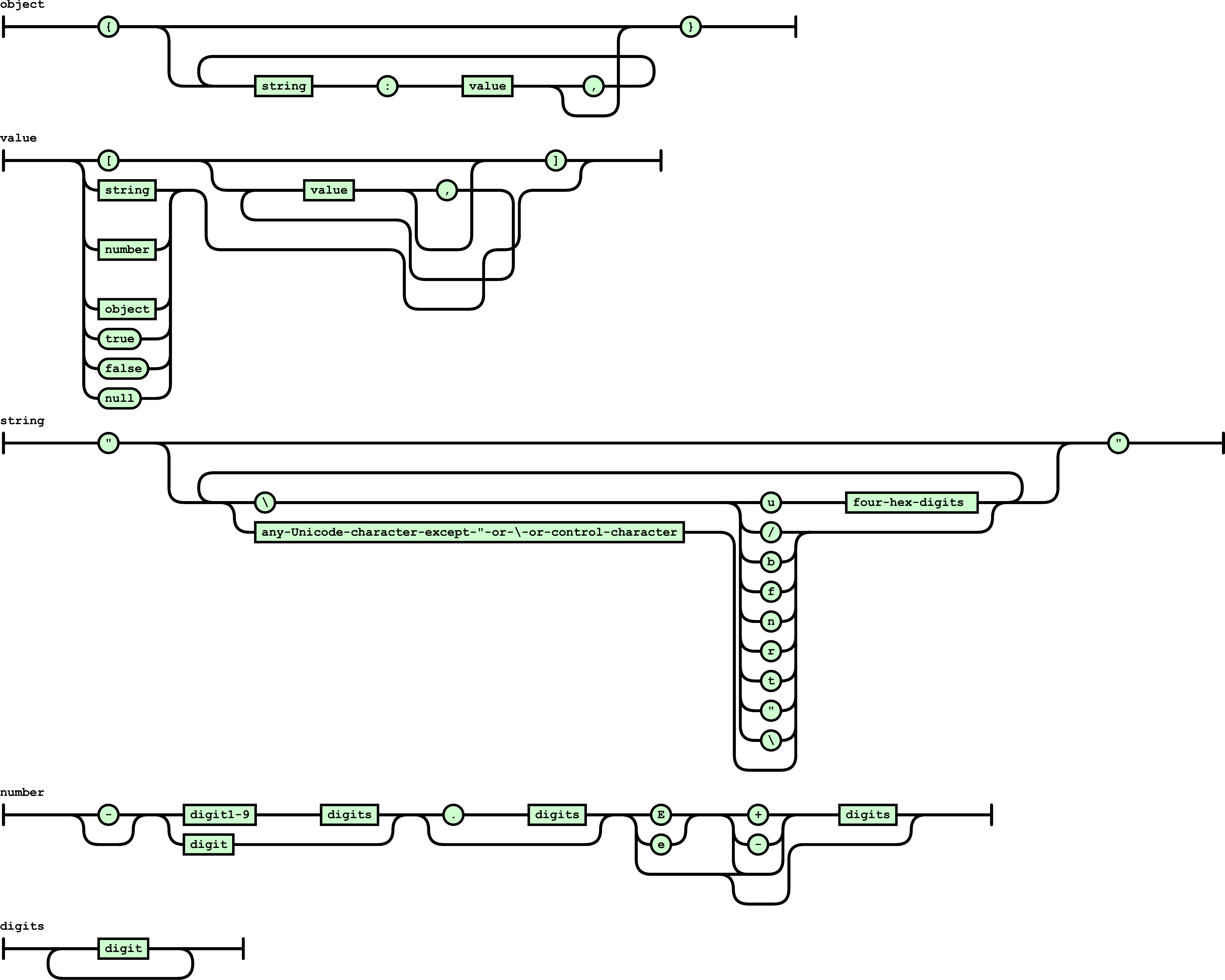}
\caption{A syntax diagram for the complete JSON grammar.}
\label{complete-json}
\end{figure}

We present in \autoref{fig:complete-lisp} and \autoref{complete-json} two complete examples of syntax diagrams of real-world grammars drawn by our implementation. For the LISP~1.5 grammar, our optimizations reduce the entire grammar to a single graph. We also present our results for the JSON grammar, which we believe (despite its obvious flaws) compares favorably with the official hand-drawn JSON syntax diagrams. Note in particular that the JSON $\nonterm{number}$ subgraph is not series-parallel, and therefore could not be represented by EBNF.

\bibliographystyle{splncs}
\bibliography{paper}

\begin{thebibliography}{10}

\bibitem{pascal-book}
Jensen, K., Wirth, N.:
\newblock {PASCAL User Manual and Report}.
\newblock Springer (1974)

\bibitem{cande-book}
Burroughs Corporation:
\newblock {Command and Edit (CANDE) Language Information Manual}. (1971)

\bibitem{lisp-book}
McCarthy, J.:
\newblock {LISP 1.5 programmer's manual}.
\newblock MIT Press (1965)

\bibitem{Bra-SJCD-1990}
Braz, L.M.:
\newblock Visual syntax diagrams for programming language statements.
\newblock SIGDOC Asterisk J. Comput. Doc. \textbf{14} (1990)  23--27

\bibitem{BelGil-Bull-1974}
Bell, S., Gilbert, E.J.:
\newblock Learning recursion with syntax diagrams.
\newblock SIGCSE Bull. \textbf{6} (1974)  44--45

\bibitem{smalltalk-book}
Goldberg, A., Robson, D.:
\newblock Smalltalk-80: The Language and Its Implementation.
\newblock Addison-Wesley Longman Publishing Co., Inc., Boston, MA, USA (1983)

\bibitem{json-web}
Crockford, D.:
\newblock {Introducing JSON}.
\newblock \url{http://json.org} (2015) Accessed: Jun 04, 2015.

\bibitem{css-web}
Atkins, Jr., T., Sapin, S.:
\newblock {CSS Syntax Module Level 3}.
\newblock \url{http://www.w3.org/TR/css-syntax-3} (2015) Accessed: Jun 04,
  2015.

\bibitem{sw1-web}
Dopler, M., Sch{\"o}rgenhumer, S.:
\newblock {EBNF Visualizer}.
\newblock \url{http://dotnet.jku.at/applications/Visualizer} (2015) Accessed:
  Jun 04, 2015.

\bibitem{sw2-web}
Thiemann, P.:
\newblock {Ebnf2ps: Peter's Syntax Diagram Drawing Tool}.
\newblock
  \url{http://www2.informatik.uni-freiburg.de/~thiemann/haskell/ebnf2ps} (2015)
  Accessed: Jun 04, 2015.

\bibitem{sw3-web}
Rademacher, G.:
\newblock {Railroad Diagram Generator}.
\newblock \url{http://bottlecaps.de/rr/ui} (2015) Accessed: Jun 04, 2015.

\bibitem{DicEppGooMen-GD-04}
Dickerson, M., Eppstein, D., Goodrich, M.T., Meng, J.Y.:
\newblock Confluent drawings: Visualizing non-planar diagrams in a planar way.
\newblock In Liotta, G., ed.: Proc. 11th Int. Symp. Graph Drawing (GD 2003).
  Volume 2912 of Lect. Notes in Comput. Sci.
\newblock Springer (2004)  1--12

\bibitem{SugTagSho-SMC-1981}
Sugiyama, K., Tagawa, S., Toda, M.:
\newblock Methods for visual understanding of hierarchical system structures.
\newblock IEEE Trans. Systems Man Cybernet. \textbf{11} (1981)  109--125

\bibitem{BasMat-DGMM-01}
Bastert, O., Matuszewski, C.:
\newblock {Layered drawings of digraphs}.
\newblock In Kaufmann, M., Wagner, D., eds.: Drawing Graphs: Methods and
  Models. Volume 2025 of Lect. Notes in Comput. Sci.
\newblock Springer (2001)  87{--}120

\bibitem{BekKauKob-JGAA-13}
Bekos, M.A., Kaufmann, M., Kobourov, S.G., Symvonis, A.:
\newblock {Smooth orthogonal layouts}.
\newblock Journal of Graph Algorithms and Applications \textbf{17} (2013)
  575{--}595

\bibitem{EppGooMen-GD-05}
Eppstein, D., Goodrich, M.T., Meng, J.Y.:
\newblock Confluent layered drawings.
\newblock In Pach, J., ed.: Proc. 12th Int. Symp. Graph Drawing (GD 2004).
  Volume 3383 of Lect. Notes in Comput. Sci.
\newblock Springer (2005)  184--194

\bibitem{HunRosSzy-JCSS-1976}
Hunt, III, H.B., Rosenkrantz, D.J., Szymanski, T.G.:
\newblock On the equivalence, containment, and covering problems for the
  regular and context-free languages.
\newblock J. Comput. Syst. Sci. \textbf{12} (1976)  222{--}268

\bibitem{StoMey-STOC-1973}
Stockmeyer, L.J., Meyer, A.R.:
\newblock Word problems requiring exponential time.
\newblock In: Proc. 5th ACM Symp. on Theory of Computing (STOC '73). (1973)
  1--9

\bibitem{GraSch-JCSS--2007}
Gramlich, G., Schnitger, G.:
\newblock Minimizing nfa's and regular expressions.
\newblock J. Comput. Syst. Sci. \textbf{73} (2007)  908--923

\bibitem{ChaCou-TCS-2004}
Champarnaud, J.M., Coulon, F.:
\newblock {NFA} reduction algorithms by means of regular inequalities.
\newblock Theor. Comput. Sci. \textbf{327} (2004)  241--253

\bibitem{HanWoo-TCS-2007}
Han, Y.S., Wood, D.:
\newblock Obtaining shorter regular expressions from finite-state automata.
\newblock Theor. Comput. Sci. \textbf{370} (2007)  110--120

\end{thebibliography}

\end{document}